\documentclass[twocolumn,prb]{revtex4}

\usepackage{graphicx}
\usepackage{amsmath}

\begin{document} 
\title{Tasting edge effects}
\author{Lyd\'eric Bocquet}
\email{lyderic.bocquet@univ-lyon1.fr}
\affiliation{Laboratoire de Physique de Physique de la Mati\`ere Condens\'ee
et Nanostructures, Universit\'e Lyon 1,
UMR CNRS 5586, 43, Bd du 11 Novembre 1918, 69622 Villeurbanne France}

%\date{\today}

\begin{abstract}
We show that the baking of potato wedges constitutes a crunchy
example of edge effects, which are usually demonstrated in electrostatics.
A simple model of the diffusive transport of water vapor around the potato
wedges shows that the water vapor flux diverges at the sharp edges in 
analogy with its electrostatic counterpart. This increased evaporation at
the edges leads to the crispy taste of these parts of the potatoes.
\end{abstract}

\maketitle

\section{Introduction}
Edge effects are usually introduced in electrostatic courses and provide an
interesting and nontrivial example of electrostatic effects. This phenomenon
corresponds to the divergence of the electric field and charge accumulation
at the edges or corners of a conductor at a fixed potential. This singular
behavior has various consequences and applications such as lightning rods
and the field emission effect.

The mathematical description of edge effect involves the solution
of the Laplace equation for the electric potential with fixed potential
boundary conditions on the conductor. For example, for a corner with an
opening angle $\alpha$ in a two-dimensional geometry, the electric field
$E$ and surface charge on the conductor behaves as $E\propto \rho^{\gamma}$
with
$\gamma=(\alpha-\pi)/(2\pi-\alpha)$ and $\rho$ the distance to the
tip.\cite{Jackson} The electric field thus diverges at the corners when
$\alpha<\pi$.

Alternative examples of edge effects can be found in other domains
of physics. The minimal ingredients are a geometry with sharp
edges; a Laplace-like equation for the physical quantity of interest
(for example, the electric potential); and a boundary condition on the given
geometry that imposes a fixed value of this quantity at its surface.

Evaporation of water vapor is one such example, as we will discuss in the
following. Edge effects arise in the context of molecular diffusion of 
water vapor. A diverging water vapor flux at the edges is predicted. Such
effects have been shown to be responsible for the formation of ring stains
formed by drying coffee drops.\cite{Deegan}
\begin{figure}[h]
\begin{center}
\includegraphics[width=5cm,height=!]{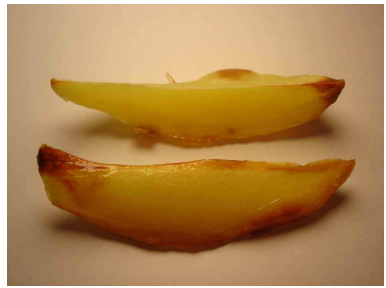} 
\caption{Two (home-made) baked potato wedges. The edges
are seen to be darker, showing dehydration in these regions.
The effect is stronger at the extremities of the wedge.} 
\label{fig1}
\end{center}
\end{figure}

We consider another example of an edge effect induced by evaporation: the
drying of potato wedges baked in an oven. The geometry of the
potatoes is fixed by the cook: we shall focus here on potatoes cut with
sharp edges, as for potato wedges. We show in Fig.~\ref{fig1} an example
of (home-made) potato wedges (after 20 minutes in the oven at
$200^\circ$C). As can be seen the edges are much darker, and exhibit very
strong drying.

We will demonstrate that this drying is due to a diverging flux of water
vapor at the extremities of the potato wedge in analogy with the edge
effect in electrostatics. This divergence induces a strong dehydration of
the potatoes in its wedges and corners. 

\section{Water vapor diffusion around potatoes}
Like most foods, potatoes contain a large amount of water. While
increasing the temperature in the oven, the liquid-vapor thermodynamic
equilibrium of water is displaced toward the vapor phase, which leads to the
evaporation of the liquid water inside the potatoes to the surrounding air.

Let us consider the distribution of water vapor in the air around potatoes.
Its concentration $c_{\rm w}$ obeys a diffusion equation:
\begin{equation}
{\partial c_{\rm w} \over \partial t} = D_{v} \nabla^2 c_{\rm w},
\end{equation}
with $ D_{v}$ the diffusion coefficient of the water vapor in air.
We make the further assumption that the oven is in a quasi-stationary
state, which implies that
\begin{equation}
\nabla^2 c_{\rm w}=0
\label{Laplace}
\end{equation}

On the potato wedge surface, the value of the vapor concentration is fixed by the liquid-vapor 
thermodynamic equilibrium, and equals the saturation value
$c_{\rm sat}$ calculated at the temperature of the oven. 
Far from the potato, we expect that the air in the oven is not saturated and
the concentration of vapor reaches a fixed value $c_\infty$ lower than
saturation. Because we are interested in the diffusion of water in the
vicinity of the potato's surface, this far field boundary condition will
not be required to characterize edge effects. The rate of evaporation of
water at the potato's surface is given by the flux of water vapor,
$J_D=-D_{v}\nabla c_{\rm w}$. %[xx what does the subscript v on D mean? xx] : Lyderic : it stands
% for 'vapor'

The equation for the vapor concentration with the specified boundary
conditions is identical to that of the electrostatic potential around a
conductor at fixed potential. The solution of Eq.~\eqref{Laplace} depends
only on the geometry of the conductor/potato's
surface.\cite{Jackson,Deegan} The vapor flux is analogous to the electric
field and is therefore expected to diverge at the edges. Both the edges and
extremities of a potato will be considered: these will be modeled
respectively in terms of two- and three-dimensional geometries. We now
recall the basic steps in deriving the solution of Eq.~\eqref{Laplace} in
the context of our problem. A detailed description can be found in
Ref.~\onlinecite{Jackson}. 
\begin{figure}[h]
\begin{center}
\includegraphics[width=5cm,height=!]{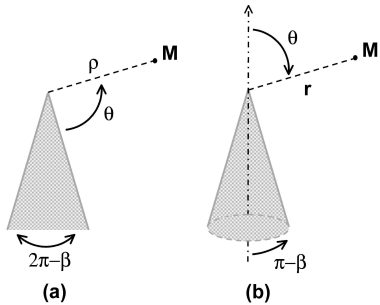} 
\caption{Geometry of the system: (a) two-dimensional wedge; (b)
three-dimensional cone.}
\label{fig2}
\end{center}
\end{figure}

\section{Drying of wedges: Two-dimensional geometry }
First consider a two-dimensional geometry as in Fig.~\ref{fig2}(a). This
geometry corresponds to evaporation at the border of a potato's wedge. The
Laplace equation for the vapor concentration is
\begin{equation}
{1\over \rho}{\partial \over \partial \rho}\Big(\rho {\partial c_{\rm
w}\over
\partial \rho} \Big) + {1\over \rho^2}{\partial^2 c_{\rm w} \over
\partial \theta^2}=0.
\end{equation}
We look for a solution using separation of variables, $c_{\rm
w}=f(\rho)g(\theta)$, and find the equations for $f$
and $g$:
\begin{subequations}
\begin{align}
{\rho\over f}{\partial \over \partial \rho}\Big(\rho {\partial f\over
\partial \rho} \Big) & =K^2 \\ {1\over g}{\partial^2 g \over
\partial \theta^2} & =-K^2,
\end{align}
\end{subequations}
with solutions $ f(\rho)=a_K \rho^{K}+ b_K \rho^{-K}$, $g(\theta)= A_K
\sin(K\theta) + B_K \cos(K\theta)$ for $K\ne 0$ and $ f(\rho)=a_0+ b_0 \log
\rho$, $g(\theta)= A_0+ B_0 \theta$ for $K=0$. 

The constant $K$ is fixed by the boundary conditions on the wedge's
surface.
To avoid a divergence of the concentration at the surface, we need to
impose
$b_0=b_K=0$. Imposing $c_{\rm w}=c_{\rm sat}$ for $\theta=0$ and
$\theta=\beta$ ($\rho>0$) yields $a_0=c_{\rm sat}$, $B_0=B_K=0$.
This condition also requires that $K$ satisfy
$\sin(K \beta)=0$, so that $K=n\pi/\beta$, with $n=1, 2,
\dots$. We put these results together
and write the vapor concentration as
\begin{equation}
c_{\rm w}(\rho, \theta)=c_{\rm sat}+\sum_{n=1}^{\infty} a_n
\rho^{n\pi/\beta} \sin \Big({n\pi \theta\over \beta}\Big).
\end{equation}
The coefficients $a_n$ are determined from the concentration far from the
corner, and it is not necessary to determine their value for the discussion
of the edge effects..

The water flux 
at the surface of the potato can now be determined using $J_D=-D_{v}
\nabla c_{\rm w}$. Only the component of the flux perpendicular to the surface
is non-vanishing. For 
$\theta=0$, we obtain:
\begin{equation}
J_D= - D_{v} \sum_{n=1}^{\infty} a_n {n\pi\over \beta} \rho^{n\pi/\beta-1}.
\end{equation}
The first term in the sum is dominant close to the wedge, and the water
flux behaves as
\begin{equation}
J_D \approx -D_{v} {a_1\pi \over \beta} \rho^{\pi/\beta -1}.
\end{equation}
As expected, this flux diverges at the edge for $\beta > \pi$, that is, 
for a sharp wedge.
The exponent of the divergence $\pi/\beta -1$ takes its maximum value,
$-1/2$, in the very sharp limit, $\beta\rightarrow 2\pi$. Note that the
same exponent ($1/2$) is found for the divergence of the surface charge
close to the edge of a thin disk at fixed potential. \cite{Jackson}

\section{Drying of edges: Three-dimensional geometry}

Our derivation can be generalized to the conical shape shown in
Fig.~\ref{fig2}(b).\cite{footnote} If we use spherical coordinates and
assume azimutal symetry, the Laplace equation becomes
\begin{equation}
{1\over r}{\partial^2 r c_{\rm w} \over \partial r^2} + {1\over
r^2
\sin \theta} {\partial \over \partial \theta} \Big( \sin\theta {\partial
c_{\rm w} \over \partial \theta}\Big)=0.
\end{equation}
We look for a solution of the form
$c_{\rm w}={f(r)\over r} g(\theta)$ and obtain the following equations for
$f(r)$ and $g(\theta)$:
\begin{subequations}
\begin{align}
{\partial^2 \over \partial r^2}f - K^{\prime} {f\over r^2}& =0 \\
{1\over \sin \theta}{\partial \over \partial \theta} \Big( \sin\theta
{\partial g \over \partial \theta}\Big)+K^{\prime} g & =0.
\end{align}
\end{subequations}
It is convenient to rewrite the unknown constant
$K^{\prime}$ as $\ell(\ell+1)$. The solution for $f$ can be written as $f(r)=
a r^{\ell+1}+b r^{-\ell}$; the solution for $g(\theta)$ is the Legendre
function of the first kind of order $\ell$,
$P_\ell(\cos\theta)$.\cite{Stegun} The regularity of the solution at the
origin imposes $b=0$. The boundary condition at the
surface of the potato, 
$c_{\rm w}=c_{\rm sat}$, leads to the following condition for the index
$\ell$:
\begin{equation}
\label{eq:this}
P_\ell(\cos\beta)=0.
\end{equation}
There is an infinite number of solutions for Eq.~\eqref{eq:this},
which we denote as $\ell_n$, with $n=1,2, \dots$.
Following the same steps as for the two-dimensional case, we obtain the
general solution as a linear combination of
solutions:
\begin{equation}
c_{\rm w}(r, \theta)=c_{\rm sat}+\sum_{n=1}^{\infty} a_n
r^{\ell_n} P_{\ell_n} (\cos \theta).
\end{equation}
The water flux on the surface thus takes the form:
\begin{equation}
J_D= D_{v} \sum_{n=1}^{\infty} a_n r^{\ell_n-1} \sin\beta
P^\prime_{\ell_n}(\cos\beta).
\end{equation}
Close to the extremity of the cone, the first term in the sum is dominant
leading to 
\begin{equation}
J_D \approx D_{v} {a_1} \sin\beta P^\prime_{\ell_n}(\cos\beta)
r^{\ell_1 -1}.
\end{equation}
\begin{figure}[h]
\begin{center}
\includegraphics[width=8.5cm,height=6cm]{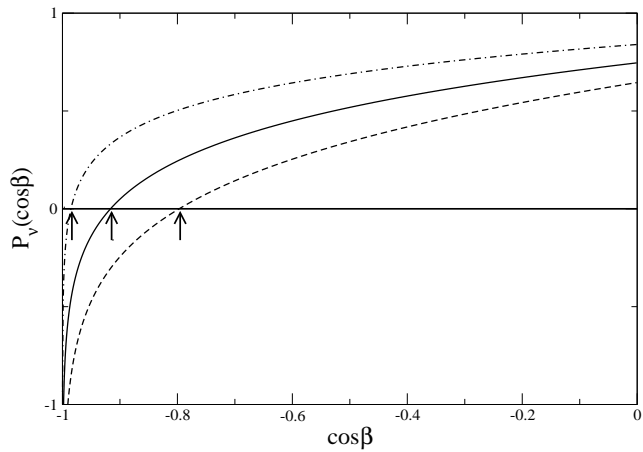} 
\caption{Plot of the Legendre function $P_\nu(\cos\beta)$ versus $\beta$ for
$\nu=0.2$, $\nu=0.3$, and 
$\nu=0.4$ (from left to right). The arrows indicate the values of
$\cos\beta$ corresponding to $P_\nu(\cos\beta)=0$. These values correspond to
$\beta\approx 170^\circ$, $\beta\approx 156^\circ$, and $\beta\approx
143^\circ$ (from left to right). The corresponding values of $\nu$ 
such that 
$P_\nu(\cos\beta)=0$ are $\nu\approx 0.4$, $\nu\approx 0.3$, and $\nu\approx
0.2$, respectively.}
\label{fig3}
\end{center}
\end{figure}

The exponent ($\ell_1-1$) characterizing the singularity of the behavior at
the sharp end is given by the smallest zero of the equation $P_{\ell_1}
(\cos\beta)=0$. In general, there is no analytical solution to this
equation, because $\ell_1$ is expected to be non-integer. For $\beta>\pi/2$,
that is, $\cos\beta<0$, the smallest 
solution of this equation is $\ell_1\in[0,1]$. In Fig.~\ref{fig3} we plot
$P_\nu(\cos\beta)$ versus
$\cos\beta$ for various $\nu$. For example, 
$P_{\nu=0.3}(\cos\beta)$ has a zero at $\cos\beta\simeq-0.91$, which
corresponds to $\beta \approx 156^\circ$. Hence
the solution of
$P_{\ell_1} (\cos\beta)=0$ for $\beta=156^\circ$ is $\ell_1=0.3$. 
Figure~\ref{fig3} also shows that the value
of % modif Lyderic
$\cos\beta$ verifying $P_{\nu}(\cos\beta)=0$ goes to $-1$ as $\nu$ decreases. Hence for
$\beta\rightarrow 180^\circ$, the value for $\ell_1$, which is a solution
of the equation $P_{\ell_1} (\cos\beta)=0$, goes to zero: $\ell_1\rightarrow
0$ as $\beta \rightarrow 180^\circ$. This result can be shown more
rigorously. In the limit $\nu\rightarrow 0$, $x\rightarrow-1$, $P_\nu(x)$
may be approximated as
$P_\nu(x)\simeq 1 + \nu \log[(1+x)/2]$.\cite{footnote2} 
%[xx $\ln$ or $\log$? xx] : Lyderic : OK, \log
Therefore, using
$\cos\beta\simeq -1+(\pi-\beta)^2/2$ as $\beta\rightarrow\pi$, we deduce
that $\ell_1$ may be approximated by
$\ell_1\simeq \Big[2\log\Big({2\over \pi-\beta}\Big)\Big]^{-1}$.
\cite{Jackson} The important conclusion of these estimates is that for a very sharp cone, $\beta\rightarrow\pi$,
the exponent of the singularity at the sharp end is $-1+\ell_1\simeq-1$ and thus
the flux divergence at the tip scales like $r^{-1}$. 

\section{discussion}

Edge effects lead to a divergent flux of the water vapor at the edges and corners of potatoes.
This singular behavior induces a strong drying of the potato near its wedges. This increased 
dehydration is responsible for the crunchy taste of the potatoes at their
wedges.

A few cooking remarks are in order. We have shown that the water flux
divergence, and thus dehydration, is stronger as the angle of
the corner or cone becomes smaller. The singularity of the water flux
is stronger for a cone shape than for a edge: the water flux scales at most
as $J_D \sim r^{-1/2}$ for the two-dimensional wedge (with $r$ the distance
to the tip), and
$J_D\sim r^{-1}$ for a sharp cone. Hence the drying of the potatoes is
predicted to be much stronger at the sharp extremities of the wedge than at
its edges as is observed (see Fig.~\ref{fig1}).

Deep fried wedges exhibit a similar
behavior with stronger dehydration at the edges. This similarity may
originate in the water vapor bubbles created at the surface of the wedge in
hot oil. The above description may apply within the bubbles, although
we expect that non-stationary effects of the diffusion process cannot be 
neglected in such a situation. 

Our simple calculations provide an interesting application of
the diffusion equation.
It has been successful in gaining the interest of the students
in a course on the physics of continuum media. Before presenting the
calculation in a lecture, students may be asked to perform their own
experiment at home and brainstorm on their observations and 
conclusions. The results may then be debated during the next lecture.

\begin{acknowledgements}
I thank Armand Ajdari for his comments on the manuscript.
\end{acknowledgements}


\begin{thebibliography}{}
\bibitem{footnote} Note that
the definition of $\beta$ has changed from that in the 2D case.
\bibitem{footnote2}This relation can be deduced by integrating 
over $\nu$ the equality ${\partial P_\nu(\cos\theta)\over
\partial\nu}\vert_{\nu=0}=2\log [\cos(\theta/2)]$ (see Eq. (8.6.20) of
Ref.~\onlinecite{Stegun}), and using $P_0(x)=1$.
\end{thebibliography}
\end{document}